# How do Chinese cities grow? A distribution dynamics approach

Jian-Xin Wu[a], Ling-Yun He[a, b,*]

[a] School of Economics, Jinan University, Guangzhou 510632, China

[b] College of Economics and Management, Nanjing University of Information Science and

Technology, Nanjing 210044, China

**Abstract**: this paper examines the dynamic behavior of city size using a distribution dynamics approach with Chinese city data for the period 1984-2010. Instead of convergence, divergence or paralleled growth, multimodality and persistence are the dominant characteristics in the distribution dynamics of Chinese prefectural cities. Moreover, initial city size matters, initially small and medium-sized cities exhibit strong tendency of convergence, while large cities show significant persistence and multimodality in the sample period. Examination on the regional city groups shows that locational fundamentals have important impact on the distribution dynamics of city size.

[*] Corresponding author. Dr. Ling-Yun He is a full professor in applied economics in the School of Economics at Jinan University, and an affiliate professor in the College of Economics and Management at Nanjing University of Information Science and Technology, email: lyhe@amss.ac.cn. Dr. Jian-Xin Wu is an associate Professor in applied economics in the School of Economics at Jinan University, twujianxin@jnu.edu.cn. This work was supported by the China National Social Science Foundation (No. 15ZDA054), Humanities and Social Sciences of Ministry of Education Planning Fund (16YJA790050) and the National Natural Science Foundation of China (Nos. 71473105, 71273261 and 71573258).

# 1. Introduction

The issue of city size distribution has received considerable attention from urban economist. Most of early works focused on testing and explaining the Zipf's law or the rank-size rule [1]. The Zipf's law states that the city size distribution follows a Pareto distribution with exponent one [2]. However, Zipf's Law can only explain the rank of large cities, but has very limited explanatory power for the distribution of small- and medium-sized cities. Therefore, recent studies tried to test whether city size distribution follows a Pareto distribution, $q$-exponential distribution, double Pareto lognormal distribution, or other distributions without restrictions on city size.

In relation with the empirical literature on the rank-size rule, a considerable number of theoretical models have been developed to shed light on the dynamics of urban growth process. The random growth model, the parallel growth models and the sequential growth models are among the most important models of them. However, both theoretic and empirical literature shows that city growth patterns may be complicated. As indicated by the sequential growth theories, the growth rate of cities may differ from city size or time. City growth patterns may be random, convergent, divergent, parallel or even hybrid [3, 4]. In the hybrid case, for example, cities may exhibit parallel growth in the long-run, but random growth in the short-run due to exogenous shocks [3, 4].

However, there are some limitations in existing literature on the distribution of city size. First, the existing literature on the evolution of city size distribution has mainly focused on explaining rank-size rule, paying little attention to other interesting aspects of the growth process. In particular, not much research has been devoted to the transitional dynamics of a system of cities towards its steady state. The rank-size rule is an abstract of real city size; however, this simplification may conceal some important information carried in the real city size. Second, most exiting studies in current literature use parametric approaches in their analysis. As indicated by Quah [5], parametric approaches only provide a statistical mean coefficient for a representative sample. This may neglect the entire shape of city size distribution and changes in the shape. Moreover, the distribution should include all sized cities, but not just the cities in the upper tail. A more useful way to explain the dynamic behavior of city size is to use a continuous distribution dynamics approach. Third, most studies use data from developed countries; few of them focus on the size distribution dynamics of cities in developing countries.

This paper focuses on the distribution dynamics of city size in China. We make the following contributions in this paper: first, instead of rank size of upper tail cities, this paper examines the cities of all sizes, providing more dynamic information about the city size, and in particular, about the entire shape of city size distribution and changes in the shape over time.

Second, this paper adopts a continuous distribution dynamics approach which can provide the dynamic law of the entire shape of city size. The distribution dynamics approach can provide full information for the transitional dynamics and long-run steady distribution of all sized cities. Several studies, such as Eaton and Eckstein [6], Anderson and Ge [7], and Xu and Zhu [8], employ discrete Markov chain approaches in their examination on the transitional dynamics of city size distribution. Compared with Markov chain approach, our continuous distribution dynamics approach requires no arbitrary discretization on the sample. Moreover, this approach can provide the full information of the transition dynamics of the entire shape of city size distribution.

Third, this paper uses Chinese city data to examine the distribution dynamics of city size. China is experiencing rapid urbanization since 1980's. The urbanization rate has increased from 23.1 percent in 1984 to 49.7 percent by the end of 2010 in China. Moreover, Chinese cities experienced quite different growth trajectories during the past three decades due to governmental policy interventions. For example, eastern coastal cities experienced far more rapid growth in population than central and western cities did. Some of the northeast and western cities even loss their population in recent years. Unbalanced economic growth may have strong impact on the growth of city size. Therefore, the dynamics of city growth in China may be significantly different from that of the stable city growth in most developed countries.

We have no intention to examine which theory can predict the distribution dynamics of Chinese city size. Instead, this paper tries to shed light on the dynamic behavior of Chinese cities during the economic transition process. We found significant persistence and multimodality in the size distribution dynamics of cities in China. Initial city size has important influence on the following transitional dynamics. Initially small- and medium-sized cities converge to their own steady states, while large cities exhibit significant persistence and multimodality. Focusing only on upper tail large cities may neglect important patterns of city growth. Moreover, locational fundamentals also matter in the distribution dynamics of city size.

Examination on the distribution dynamics of city size has important policy implications. Urbanization is not just about the increasing population in urban areas, but also about the spatial distribution of population and economic activities across different cities. This spatial distribution of population and economic activities is closely related with economic efficiency. During the rapid urbanization process in recent decades, urban investment accounts for largest share of the total investment in China. Most of these investments are extremely durable, knowing the distribution dynamics may help to make reasonable urban planning.

The remainder of the paper is structured as follows: Section 2 reviews some of the related empirical and theoretical literatures. Section 3 introduces the methodological issues and data. Section 4 provides some preliminary analyses on the current city size distribution in China. The main empirical findings are provided in Section 5. The final section concludes the paper.

**2. Literature review**

The city size distribution has received much attention in economic research. In general, there are two kinds of literature on the distribution of city size. The first concerns about the shape of the city size distribution from a static perspective. The second examines how cities growth relative to each other from the dynamic perspective, i.e., the dynamic evolution of city size. In this section, we provide a brief review of these two strands of literature.

Empirically, most of the existing literature on the city size distribution has mainly focused on testing and explaining Zipf's law or the rank-size rule. However, the Zipf's Law is the special case of Pareto law when the exponent equals one [2]. Therefore, a considerable number of studies try to test if city size follows a Pareto law [7, 9-15]. Beside the traditional Pareto distribution, there are also studies on other distributions, such as the q-exponential distribution [16, 17], double Pareto lognormal distribution [18-19, 2], and the combination of lognormal and a Pareto distribution [20]. Due to differences in samples and econometric models, the empirical results are quite mixed. More empirical researches are expected to shed light on the distribution of city size.

Three strands of literature are closely related with the dynamic evolution of city size distribution. The first is the theory of random growth, which assumes that city growth process have the common expected growth rate and common standard deviation. The random growth theory predicts that the city size distribution follows Gibrat's law. Gabaix [21] establishes the relationship between Gibrat's laws and Zipf's Law by formally proving that Gibrat's law implies Zipf's Law. Córdoba [22] derives restrictions on some factors, such as preferences, technologies, and stochastic properties of the exogenous driving forces, that urban models must satisfy to explain why city size distribution follows Pareto distribution. Rossi-Hansberg and Wright [3] develop a general equilibrium model of economic growth in an urban environment. Their model produces a city size distribution that is well approximated by Zipf's law.

The second strand of literature is the theory of the parallel growth. The theory of the paralleled growth can well explain the Zipf's law. Eaton and Eckstein [6] examine the evolution of the transition matrices of France's and Japan's largest metropolitan areas. Their results show that the size-distributions in steady states similar to what they have been in the history. Instead of convergence to an optimal city size or the divergent growth of the largest cities, they conclude that city size takes the form of parallel growth pattern. Black and Henderson [11] develop an endogenous urban growth model to describe how human capital accumulation, and knowledge spillover foster the parallel growth of cities. Sharma [23] uses Indian city data and time-series approach, and finds that the growth of cities may be parallel in the long-run. However, there are may be deviations from the long-run parallel growth of cities due to exogenous shocks in the short-run. Wang and Zhu [24] find a parallel growth pattern of Chinese cities in the long run. Chen et al. [4] use Chinese city data and time-series econometric techniques to examine the parallel growth. Their results show that overall Chinese city growth does not follow parallel growth. The parallel growth only exists in some city groups with certain characteristics, such as similar location advantages or policy regime.

The third strand of literature examines the Sequential growth of city size. Henderson and Venables [25] and Cuberes [26] develop models of city formation in which urban agglomerations grow sequentially. In these models, the initially largest cities are the first to grow until they reach a critical size, at which point they are followed by the second-largest cities, then the third-largest ones, and so on. Empirically, using two comprehensive data sets on populations of cities and metropolitan areas for a large set of countries, Cuberes [27] finds strong evidence for the sequential growth of city size. Sánchez-Vidal et al. [28] use the US new-born cities in the twentieth century to examine the distribution dynamics of city size. Their findings support the sequential growth of city size. Using the US city size data, Sheng et al. [29] also find that cities grow in sequential patterns. Their results also show that city size in the upper tail exhibits significant conditional convergence characteristics.

## 3. Methodology and data

Traditional econometric techniques only provide information on the growth behavior of an average or representative economy [5]. They provide no information how cities of different sizes evolve relative to each other. They shed little light on some interesting questions on the long run distribution of city sizes. For example, is the city sizes growth parallel, diverge or converge? If they converge, is the long run steady state unimodal, bimodal or multimodal? To address these questions, one needs to know the entire cross-sectional distribution dynamics of city size in the sample.

The distribution dynamics approach developed by Quah [5, 30, 31] is widely used to examine the dynamic behavior of per capita income distribution. There are two types of distribution dynamics approaches: the discrete approach and the continuous approach. Early studies use a discrete approach to estimate transition probability matrix and the ergodic distribution. This arbitrary discretization of state space into limited state have several limitations: first, it cannot provide any mobility information within state; second, it may change the probabilistic properties of variable in question; third, the estimated results of this approach are sensitive to this discretization approach [31-33]. To overcome these limitations, Quah [31] develops a stochastic kernel approach to estimate the transition probability and ergodic distribution. This approach requires no discretisation for the variables. Therefore, it can be considered as the upgraded version of the discrete one. This paper use the continuous distribution dynamics approach to examine the dynamic behavior of city size in China.

To estimate the long run (ergodic) distributions of city size, we need to calculate the transition probability distribution first. Let $f_t(x)$ denote the distribution of variable $x$ at time $t$, and $f_{t+\tau}(y)$ denote the distribution of variable $y$ at time $t+\tau$, where $\tau>0$. Assuming that the evolution of the distribution is time-invariant and first-order, that is, the future distribution of variable $y$ in $t+\tau$ depends only on the distribution of variable $x$ in time $t$. The relationship between the two distributions, namely $f_t(x)$ and $f_{t+\tau}(y)$, can be written as follows:

$$f_{t+\tau}(y) = \int_0^\infty g_\tau(y|x) f_t(x) dx \tag{1}$$

where the conditional density term $g_\tau(y|x)$ is a transition probability operator, which is a stochastic kernel mapping the transition process of the distribution from time $t$ to time $t+\tau$. For any $x$, we have $\int_0^\infty g_\tau(y|x) dy = 1$ (See Johnson [33] for further calculation details).

In this paper, we use kernel density method to estimate the transition probability distribution and its ergodic distribution. The joint natural kernel estimator of $f_{t,t+\tau}(y,x)$ can be estimated as follows:

$$f_{t,t+\tau}(y,x) = \frac{1}{nh_xh_y}\sum_{i=1}^{n} K_x(\frac{x-x_i}{h_x})K_y(\frac{y-y_i}{h_y}) \quad (2)$$

where $h_x$ and $h_y$ denote the bandwidth of $x$ and $y$ respectively. $x_i$ and $y_i$ are the population size values of the cities at time $t$ and time $t+\tau$, respectively. Following most studies [31, 33, 34], the bandwidths in this paper are calculated as the optimal method described in Silverman [35].

Similarly, we can further define the marginal kernel of $x$ as

$$f_t(x) = \frac{1}{nh_x}\sum_{i=1}^{n} K_x(\frac{x-x_i}{h_x}) \quad (3)$$

With the joint and marginal kernel density function defined above, the conditional density can be calculated as follows:

$$g_\tau(y|x) = \frac{f_{t,t+\tau}(y,x)}{f_t(x)} \quad (4)$$

Given that the transition probability function ($g_\tau(y|x)$) remains unchanged, this transition dynamics will evolve into a long run equilibrium state, i.e. the ergodic distribution. Therefore, the ergodic distribution (denoted by $f_\infty(y)$), which corresponds to a given $g_\tau(y|x)$, given that it exists, can be estimated through:

$$f_\infty(y) = \int_0^\infty g_\tau(y|x)f_\infty(x)dx \quad (5)$$

In the continuous dynamic distribution approach, three-dimension plots and contour plots are the most often used tools to present the transition probability, while the ergodic distribution is displayed with a kernel density plot. To get more information on the transition dynamics of the distribution, this paper adopts a net transition probability approach proposed by Cheong and Wu [36]. This approach provides the precise values of net upward probability at each point. Intuitively, a positive net transition probability at a point suggests the increase of city size, while a negative net transition probability at a point implies the decrease of city size. The net transition probability $p(x)$ can be estimated as

$$p(x) = \int_x^\infty g_\tau(z|x)dz - \int_0^x g_\tau(z|x)dz \qquad (6)$$

Following common practice in distribution dynamics approach, this paper uses relative city size (RCS) which is the population size of each city divided by its yearly average in the analysis.

Chinese city data sets are collected at two levels: county level and prefectural level. From the perspective of agglomeration and administration, prefectural city tends to be more like an urban system than county level city. Due to data availability, this paper uses 237 prefectural cities for the period 1984-2010. Following existing studies (such as Anderson and Ge[7]; Chen et al.[4]), city size is defined as the number of non-agricultural population in a city (by permanent residence) at the year's end. All our data are from each year's *Chinese City Statistical Yearbook*. [37]

## 4. The preliminary analysis of city size distribution

Figure 1 shows the distribution of RCS in 1984, 1990, 2000 and 2010. We can observe that the peak of the distribution locates below mean value (RCS=1). This implies that most of Chinese cities have population size below average. In 1984, 67.5 percent of Chinese cities are distributed below average; in 2010, this number increased to 70 percent. Small sized cities are dominant in Chinese cities. The peak point of RCS shows an increase trend in our sample period, with shrinkage of probability mass at both ends of the distribution. This suggests that the size distribution of Chinese cities is more compacted in 2010 than that in 1984. Therefore, the distribution of critical years may imply convergence of city size in China. This result is consistent with the findings of Anderson and Ge [7], which shows that the distribution of city size in China remains stable before the Reform but show significant convergence during the post-reform period.

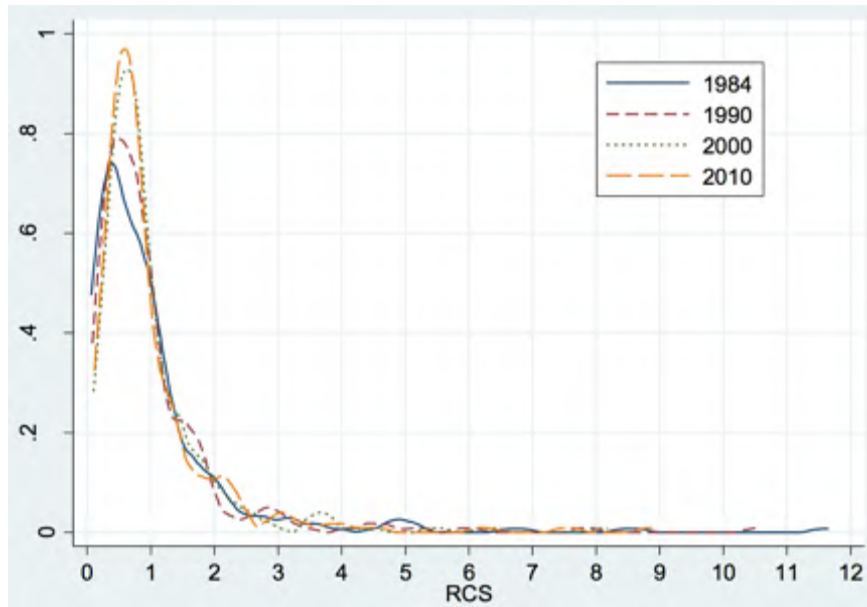

Figure 1 Kernel density distribution of all cities in representative years

The coefficient of variation is used to measure the sigma convergence or dispersion. For robustness, we estimate the coefficients of variation for city size in each year, and plot them in Figure 2. The result shows that the coefficient of variation follows a decline trend in the period 1984-2003, and an increase trend after 2003. This is broadly consistent with the distributions of RCS in the representative years. However, one may ask, what drives the structural changes of distribution in 2003? This may relate with the abolishment of custody and repatriation system (Shōu-Róng-Qiǎn-Sòng-Zhi-Du) in 2003. Besides household registration system (Hukou), Chinese government established custody and repatriation system to restrict migration between different regions in early time[39]. Chinese government adopted the policy of encouraging the development of the small and medium-sized cities while controlling the population in large cities, particularly in megacities like Beijing, Shanghai, and Guangzhou. In fact, the custody and repatriation system is far stricter in large cities than in small and medium-sized cities. Therefore, the abolishment of custody and repatriation system significantly increased the population in large cities.

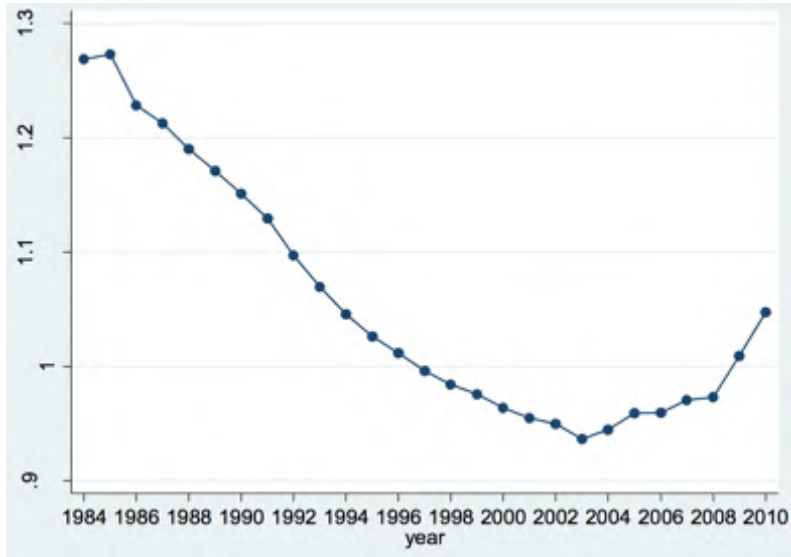

Figure 2 The coefficient of variation, 1984-2010

**5. The distribution dynamics of city size in China**

*5.1 All cities*

Figure 3 plot the distribution dynamics of all cities for the period 1984-2010. The three-dimensional plot of transition probability of RCS (see Panel (a) of Figure 3) shows the distribution of the probability mass with which a city with a specific RCS at time $t$ could evolve into each RCS value at time $t+1$. To make it more intuitively, suppose that we chose the point with a RCS value of 3 on the $t$ axis, and slice the plot from this point parallel to axis $t+1$, this slice shows the transition probability mass distribution of a city with RCS value of 3 transitioning into each value of RCS at time $t+1$. Therefore, the concentration of probability mass along the diagonal implies persistence in relative position changes among cities, while deviation from the diagonal suggests mobility in relative positions. Moreover, if the distribution of probability mass parallel to the $t$ axis, this may suggest a process of convergence. On the contrary, if the distribution of probability mass parallel to the $t+1$ axis implies divergence. The contour plot in Panel (b) of Figure 3 is a top down view of the three-dimensional graph in Panel (a), which can show more clearly the deviation from the diagonal line. The contour plots work like the contours on a standard map, connecting points at similar heights on the stochastic kernel.

The three-dimensional plot of transition probability of RCS in Panel (a) of Figure 3 and its contour plot in Panel (b) show that the probability mass is significantly situated along the diagonal. This implies that persistence is the main feature of the transitional dynamics for this period. However, the cities with RCS values greater than 4 have more transition dynamics than cities with RCS values smaller than 4. This means that large cities, particularly megacities, tend to have greater motility in their relative city size. On the other side, several probability mass peaks and discontinuity can be observed in the three-dimensional plot and contour plot, suggesting the possible existence of convergence clubs in the long run steady state.

Panel (c) of figure 3 shows the distribution of net transitional probability for RCS of Chinese cities for the period 1984-2010. We can observe that the net transition probability in high RCS end is higher than in the low RCS end. This implies that large cities tend to have more mobility than small and medium-sized cities. This result further confirms the findings showed in three-dimensional and contour plots. Moreover, no evidence of convergence in terms of city size can be found among large cities with RCS values above 4. However, in the region with RCS values below 4, there is only one cross point with 0 axis in RCS value of 0.8. Moreover, cities with RCS values above 0.8 have negative net transition probability, while cities with RCS values below 0.8 have positive net transition probability. This implies strong convergence in terms of city size across Chinese cities with RCS values smaller than 4. In general, 61.8 percent of observations have a tendency to reduce their relative size for the period 1984-2010, implying the trend of convergence.

Panel (d) of figure 3 shows the ergodic distribution for RCS of Chinese cities for the period 1984-2010. Multimodality can be observed in the long-run distribution. There are four peaks in the ergodic distribution, locate around the RCS values of 0.8, 4.8, 6.5, and 8.2 respectively. This result suggest that, keeping the transition dynamics in our sample period unchanged, Chinese cities will converge into different city size groups in the long run steady state. This finding is different from the result predicted by theories of paralleled grow cities. Differing from the findings in Eaton and Eckstein (1997), this paper finds that the city size distribution in the long run steady state significantly differ from the city size distributions in representative years as showed in previous section.

(a) Three-dimensional plot of transition probability (b) Contour plot of transition probability

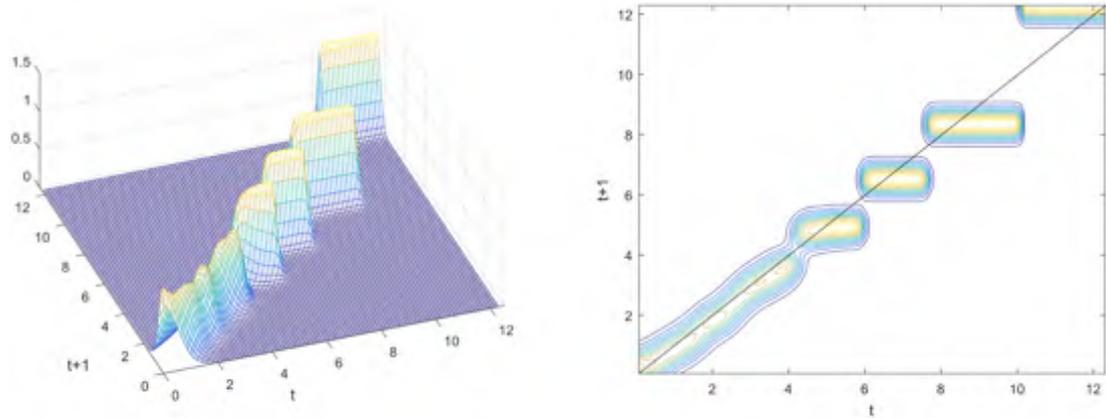

(c) Net transition probability plot      (d) Ergodic distribution

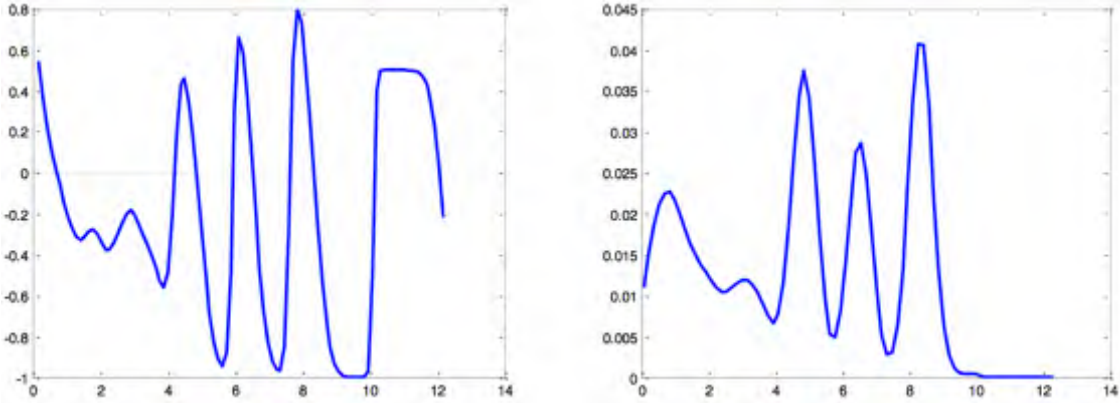

**Figure** 3 The distribution dynamics of city size for the period 1984-2010

*5.2 Distribution dynamics in sub-periods of 1984-2003 and 2003-2010*

As indicated by the results of the coefficient of variation in figure 2, the evolution trend for dispersion of city size changes significantly in 2003. To get more insight on this point, we estimate the distribution dynamics of city size for two sub-periods of 1984-2003 and 2003-2010. The results are plotted in Figures 4 and 5, respectively. The distribution dynamics for the period 1984-2003 are almost the same as those for the period 1984-2010. Compared with that for the period 1984-2003, the ergodic distribution for the period 2003-2010 is more compacted. Moreover, 59.5 percent of cities decreased their RCS values for the period 1984-2003, while 68 percent of cities decreased their RCS values for the period 2003-2010. This implies that there are more cities have a down-moving tendency in the second sub-period than in the first sub-period in terms of city size. Although there are some differences in the external shapes of the distributions between two sub-periods, persistence and multimodality are still significant in the transition probability distribution and ergodic distribution in both sub-periods.

(a) Three-dimensional plot of transition probability  (b) Contour plot of transition probability

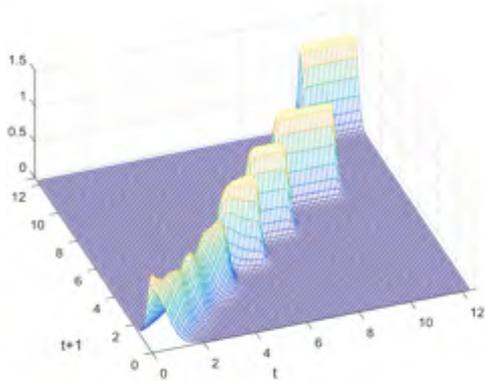 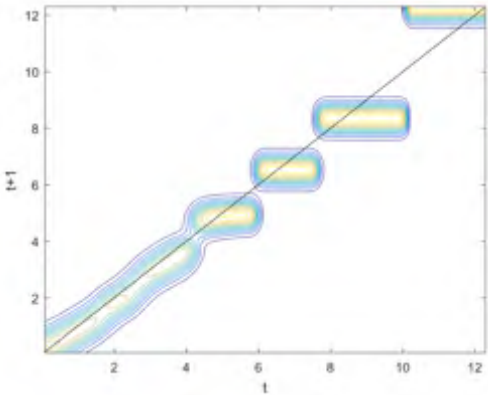

(c) Net transition probability plot  (d) Ergodic distribution

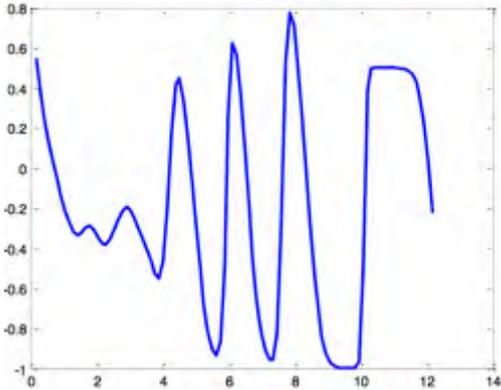 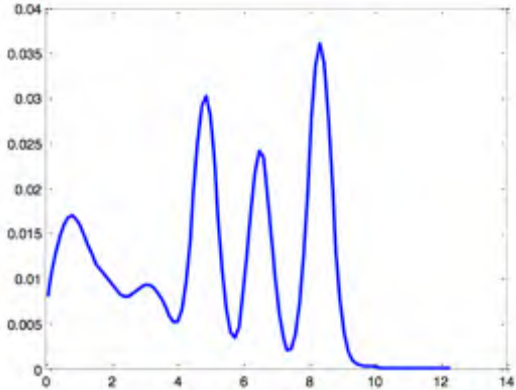

**Figure** 4 The distribution dynamics of city size for the period 1984-2003

(a) Three-dimensional plot of transition probability (b)Contour plot of transition probability

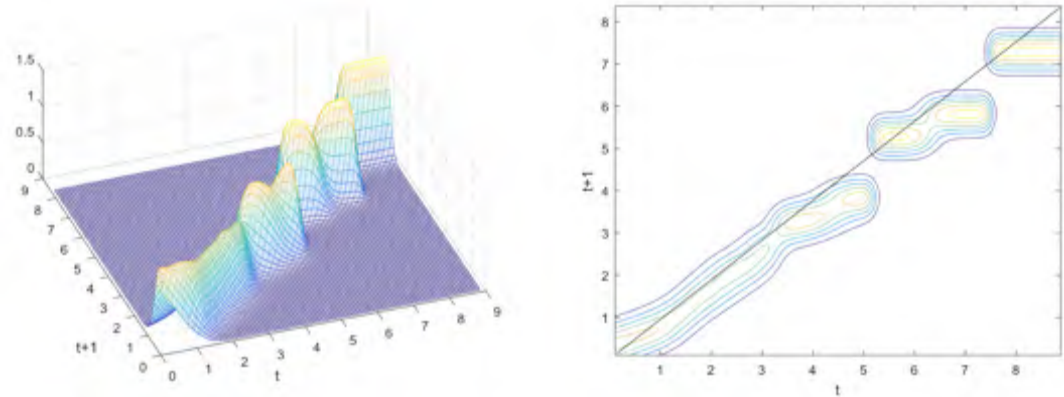

(c) Net transition probability plot (d) Ergodic distribution

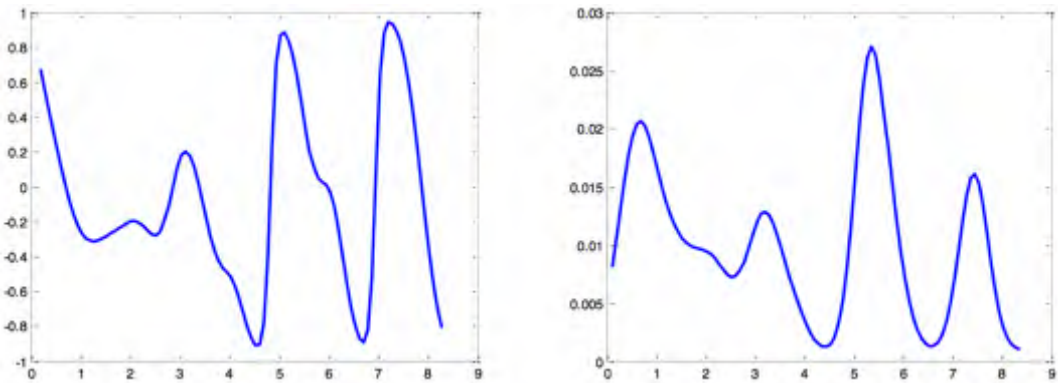

**Figure** 5 The distribution dynamics of city size for the period 2003-2010

*5.3 Distribution dynamics in different size groups*

Many studies [25, 26] have indicated that initial city size have significant impact on the following growth of cities, implying the existence of path dependence effect. To further understand how initial city size affects the following distribution dynamics, we divided all cities into three equal groups according to their population size in 1984, namely small cities (79), medium-sized cities (79) and large cities (78). Therefore, this sub-section will focus on the distribution dynamics in various size groups.

*(1) Large cities*

Figure 6 show the distribution dynamics of large cities for the period 1984-2010. Similar to the results of all cities, both the three-dimensional plot of panel (a) and the contour plot in panel

(b) show that the probability mass are distributed along the diagonal. Persistence is the major feature in the transition dynamics of size for large cities. However, there is more mobility in high RCS end than in low RCS end. The net transition probability plot in panel (c) of figure 6 shows that there are more negative regions than positive regions in the horizontal axis. This means that more cities have down-moving tendency in the intra-distribution changes. In fact, 73.2 percent of large cities in our sample have a decrease in their RCS values. The ergodic distribution in panel (d) of figure 6 shows significant multimodality. This implies that, keeping the transition dynamics remain unchanged, the long run steady distribution will exhibit club convergence.

(a) Three-dimensional plot of transition probability  (b) Contour plot of transition probability

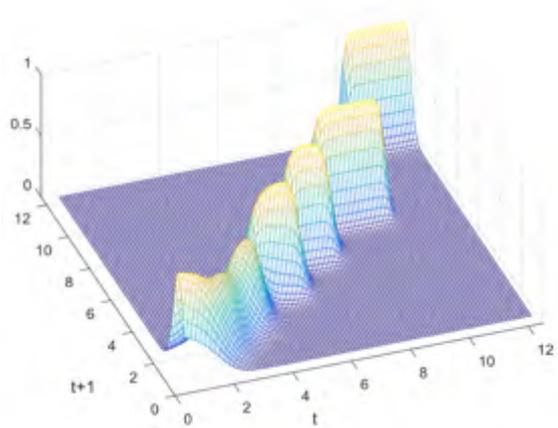 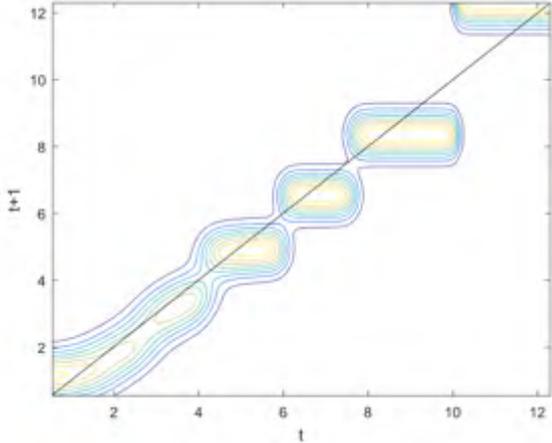

(c) Net transition probability plot  (d) Ergodic distribution

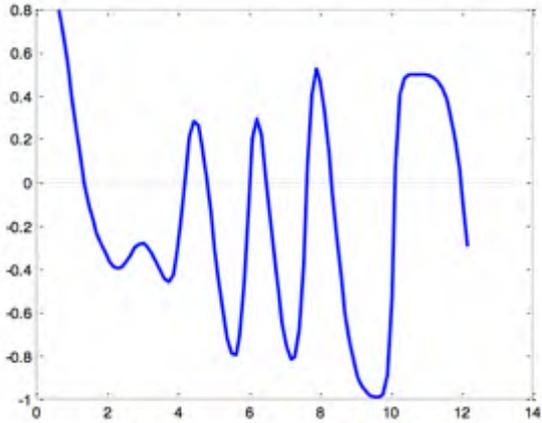 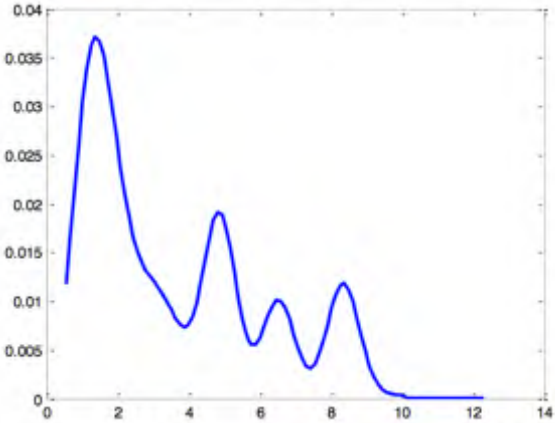

**Figure** 6 The size distribution dynamics of large cities for the period 1984-2010

*(2) Medium-sized cities*

Figure 7 show the distribution dynamics of medium-sized cities for the period 1984-2010. Compared with that of large cities, the transition probability mass of the medium-sized cities tend to have more mobility. The three-dimensional plot in panel (a) and contour plot in panel (b) of figure 7 show that the transition probability mass is almost parallel to the t axis. This may mean that medium-sized cities tend to converge into a specific city size. This result is further confirmed by the net transition probability plot in panel (c) of figure 7.

(a)Three-dimensional plot of transition probability    (b) Contour plot of transition probability

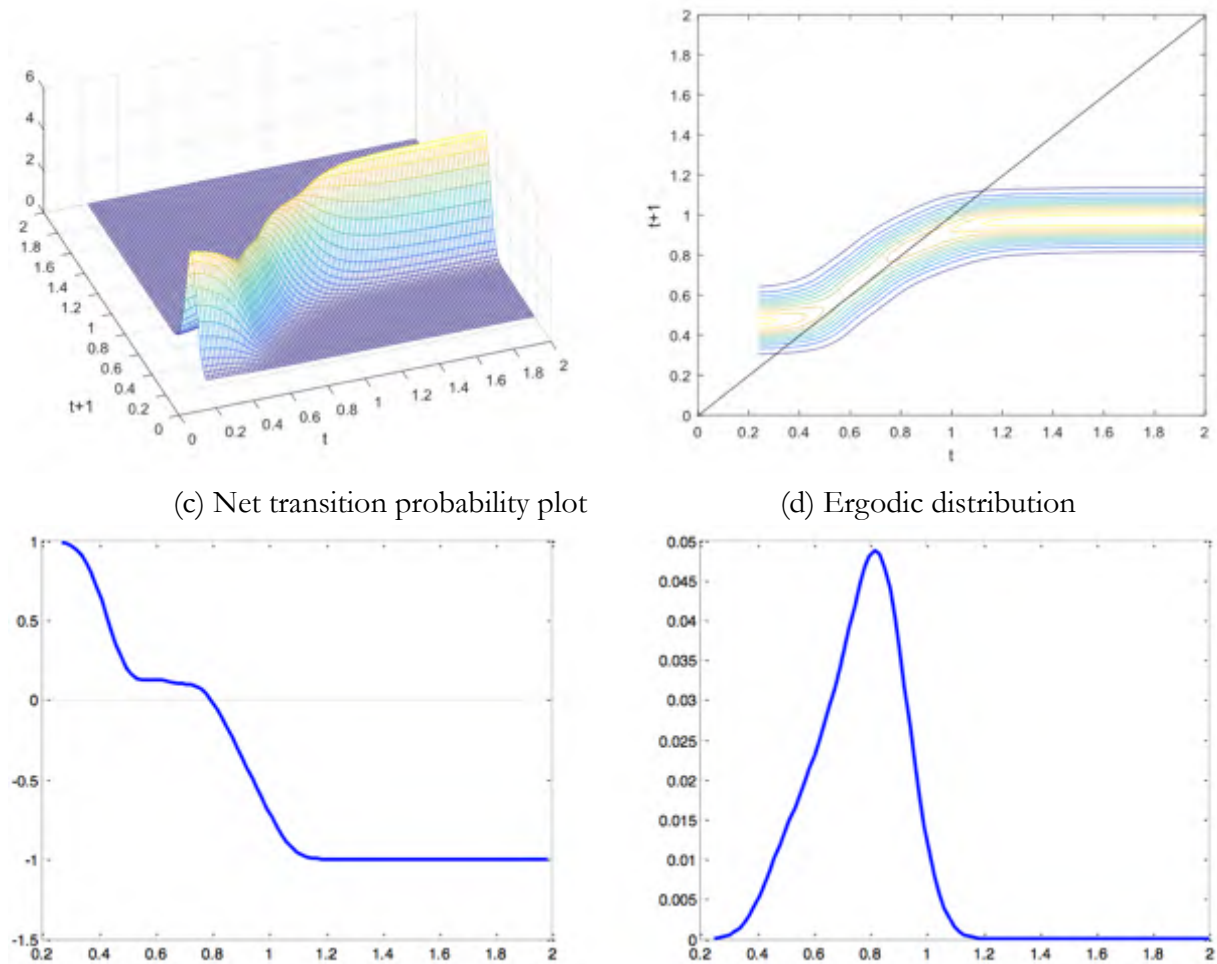

(c) Net transition probability plot    (d) Ergodic distribution

**Figure** 7 The size distribution dynamics of medium-sized cities for the period 1984-2010

The medium-sized cities in the initial year of 1984 with RCS values smaller than 0.8 have a net positive transitional probability, implying that these cities have a strong tendency to increase their relative sizes. However, the medium-sized cities with RCS greater than 0.8 have net negative transition probabilities, indicating that those cities have strong tendency to reduce their relative sizes. Correspondingly, the ergodic distribution in panel (d) of figure 7 exhibit significant uni-modality, with the peak situates around the RCS value of 0.8. This implies that, keeping this transitional dynamics unchanged, the medium-sized cities in 1984 will converge to 0.8 times average city size in the long run steady state.

*(3) Small cities*

Figure 8 show the distribution dynamics of small cities for the period 1984-2010. The three-dimensional plot in panel (a) and contour plot in panel (b) of figure 8 show that transition probability mass of cities with RCS values above 0.4 is distributed parallel to the t-axis, while the transition probability of cities with RCS values below 0.4 have relatively strong persistence. Panel (c) of figure 8 shows the distribution of net transition probability. It indicates that the cities with RCS values below 0.31 have tendency to increase their city size, while cities with RCS values above 0.31 have tendency to decrease their city size. The ergodic distribution in panel (d) of figure 8 indicates that, keeping this transition dynamics unchanged, the distribution of small cities will converge into a steady distribution around the 0.31 times average city size.

(a)Three-dimensional plot of transition probability (b) Contour plot of transition probability

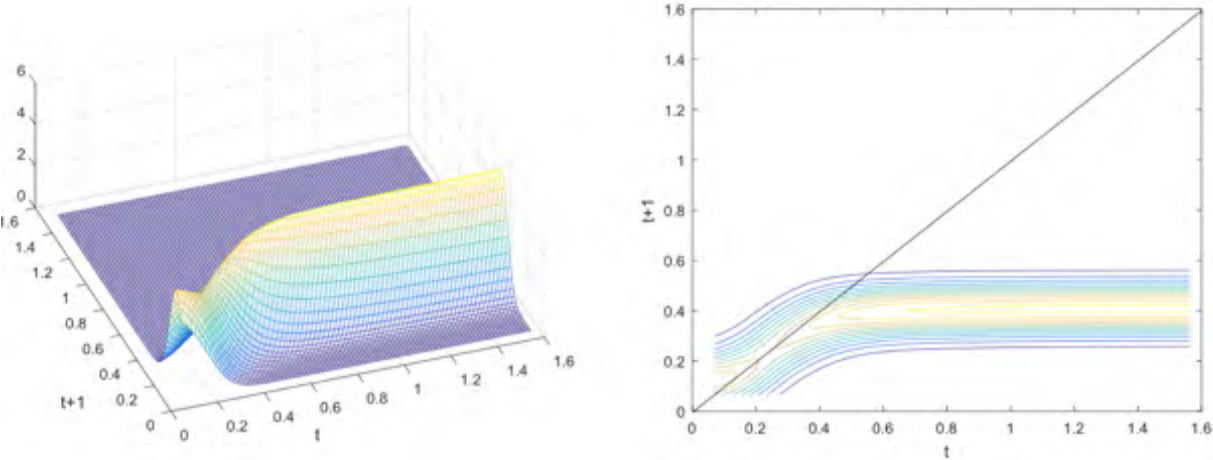

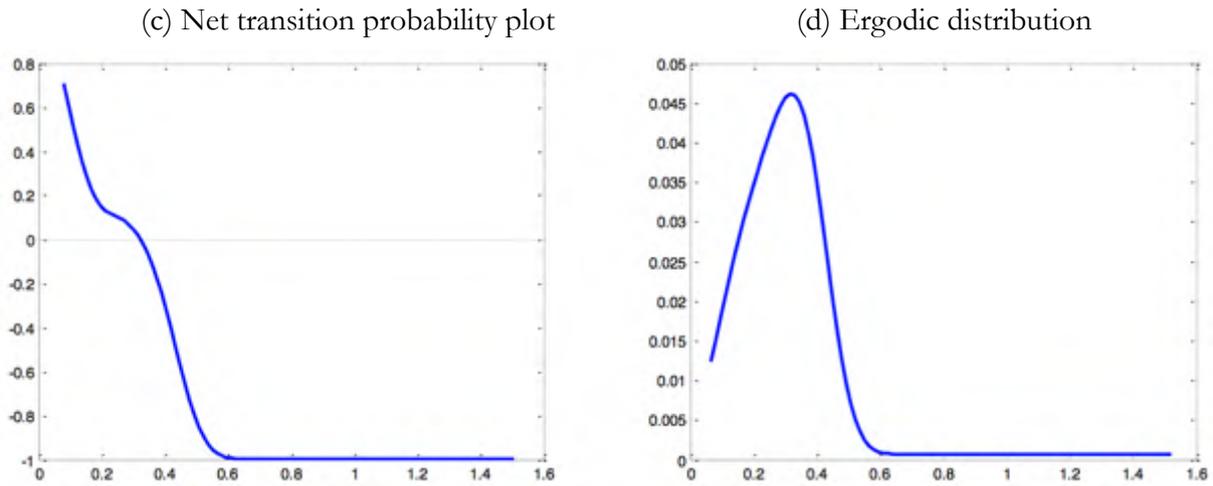

**Figure** 8 The size distribution dynamics of small cities for the period 1984-2010

*5.4 Distribution dynamics in regional groups*

Locational fundamentals may play crucial role in the formation and distribution dynamics of city size [38]. China is a large country with many heterogeneous regions, which differ greatly in terms of natural resource endowment, weather conditions, transport accessibility and government policy intervention. These factors may have strong impact on the distribution and growth of cities [4]. Therefore, in this subsection we will examine the distribution dynamics of city size in different regional groups. Following most studies, we divide the full sample of Chinese cities into three sub-datasets according to three regions, namely the eastern, central and western regions. The eastern regions include: Beijing, Tianjin, Hebei, Liaoning, Shanghai, Jiangsu, Zhejiang, Fujian, Shandong, Guangdong, Guangxi, and Hainan; The central regions include: Jilin, Heilongjiang, Inner Mongolia, Shanxi, Anhui, Jiangxi, Henan, Hubei, Hunan; The western regions include: Chongqing, Sichuan, Guizhou, Yunnan, Shaanxi, Gansu, Qinghai, Ningxia, and Xinjiang. There are 105, 84, 47 cities in the eastern, central and western regions respectively.

*(1) The eastern cities*

Figure 9 shows the distribution dynamics of the eastern cities for the period 1984-2010. Both three-dimensional plot (panel (a) of figure 9) and contour plot (panel (b) of figure 9) show that persistence is the dominant characteristics in the distribution of transition probability mass. Similar to that in all cities, cities with RCS values greater than 4 have more mobility than cities with RCS values smaller than 4. The net transition probability plot in panel (c) of figure 9 shows that there is no significant evidence of convergence among the eastern cities. Consequently, keeping the transition dynamics unchanged, the long run ergodic distribution of the eastern cities exhibit significant multimodality.

(a) Three-dimensional plot of transition probability    (b) Contour plot of transition probability

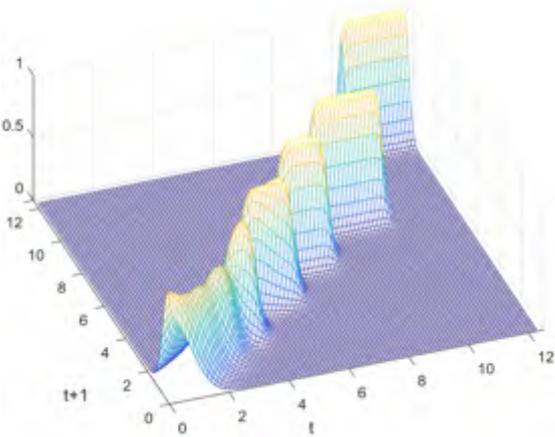 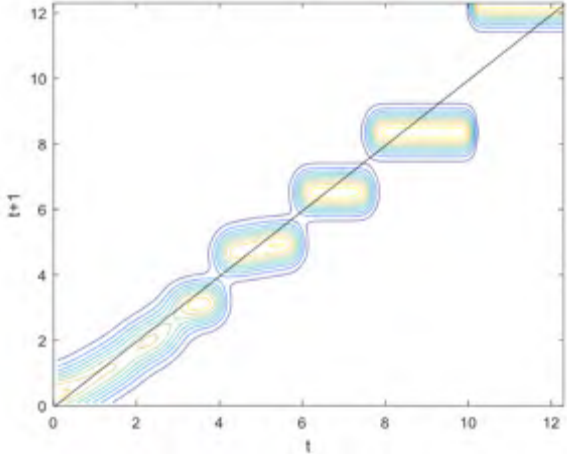

(c) Net transition probability plot    (d) Ergodic distribution

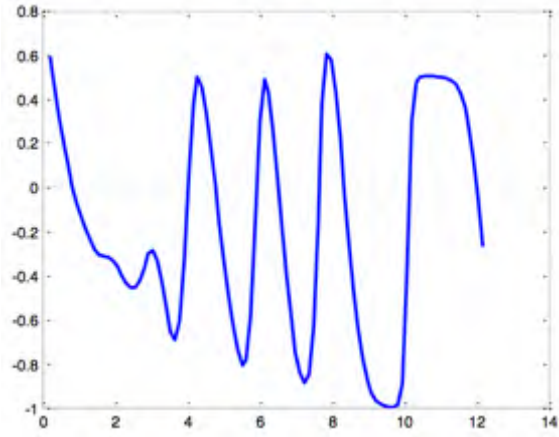 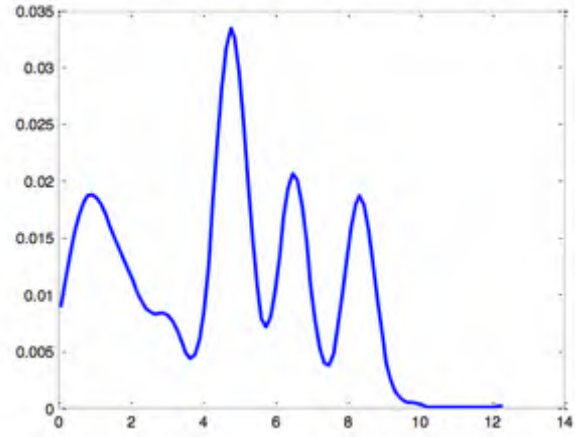

**Figure** 9 The size distribution dynamics of the eastern cities for the period 1984-2010

*(2) The central cities*

Figure 10 shows the size distribution dynamics of the central cities for the period 1984-2010.

Compared with the eastern cities, there are few megacities in the central region. Therefore, the size distribution of the central cities is much more compacted than the eastern cities. Both three-dimensional plot and contour plot show that transition probability mass situate along the diagonal, indicating that persistence is the dominant characteristics of the intra-distribution changes in city size among the central cities.

(a) Three-dimensional plot of transition probability (b) Contour plot of transition probability

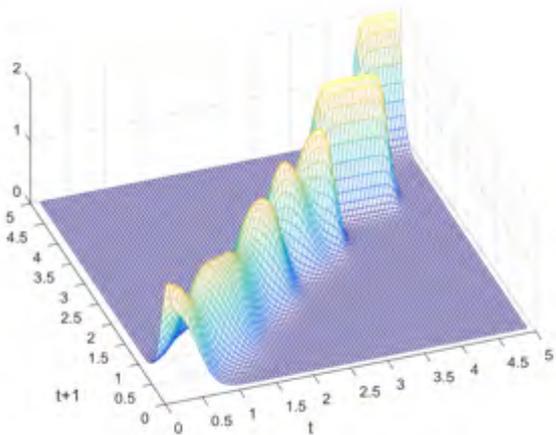 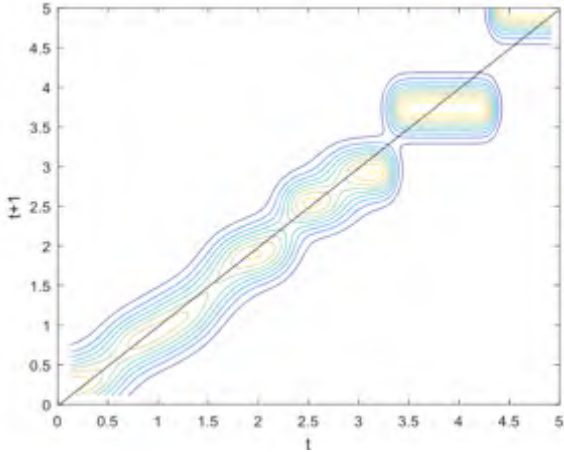

(c) Net transition probability plot (d) Ergodic distribution

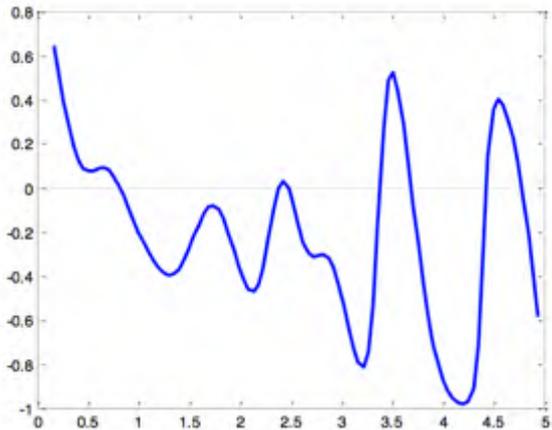 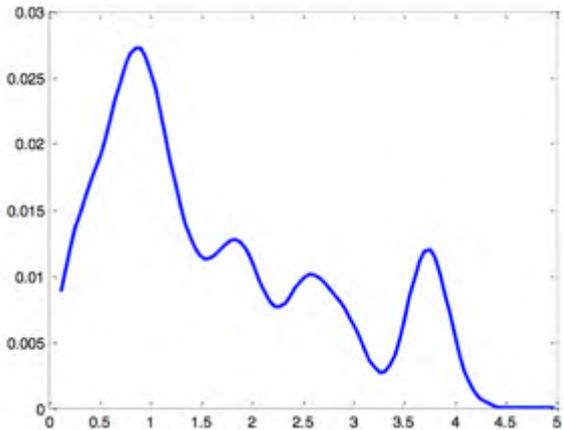

**Figure** 10 The size distribution dynamics of the central cities for the period 1984-2010

The net transition probability plot in panel (c) of figure 10 shows that there are only a few small regions have positive net transition probability, while most regions have negative net transition probability. This indicates that cities have the tendency to decrease their relative city size are more than cities tend to increase their RCS in the central region. In fact, 67.2 percent of observations have tendency to move down in their relative position changes in the central cities, which is much higher than 58.5 percent in the eastern cities and 59.4 percent in the western cities. Significant multimodality can be observed in the ergodic distribution of the central cities, implying multiple steady states also exist in the central cities group.

*(3)The western cities*

Figure 11 plots the distribution dynamics of RCS for the western cities for the period 1984-2010. The transition probability mass distribution of the western cities seems to have more mobility than that of the central and eastern cities. Apparent deviation from the diagonal can be observed in the high RCS end of the transition probability mass distribution in both panel (a) and panel (b) of figure 11, implying more city size mobility in large cities than in small and medium-sized cities. The net transition probability plot in panel (c) of figure 11 shows no significant evidence of convergence in city size among the western cities. Similar to the results for the eastern and central cities, the ergodic distribution in panel (d) of figure 11 shows that the dominant characteristics in the long run distribution is multimodality.

(a)Three-dimensional plot of transition probability    (b) Contour plot of transition probability

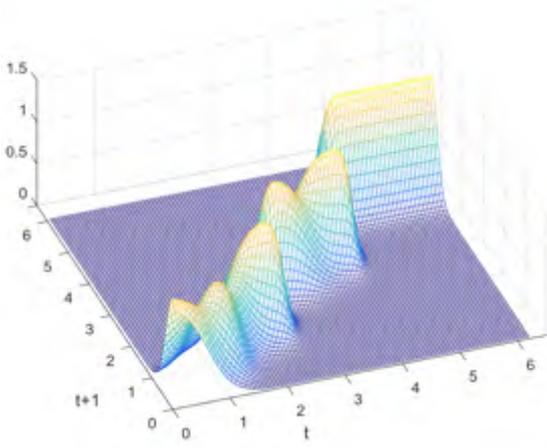 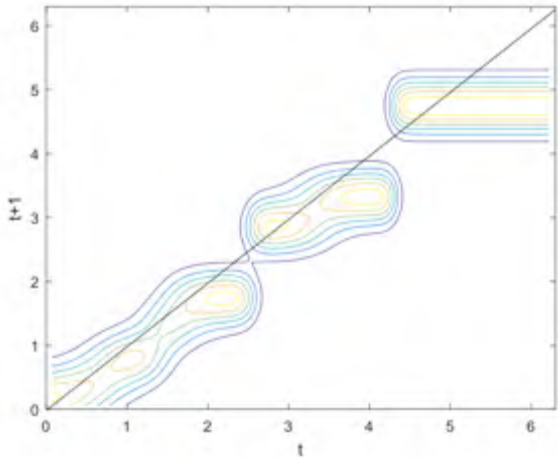

(c) Net transition probability plot    (d) Ergodic distribution

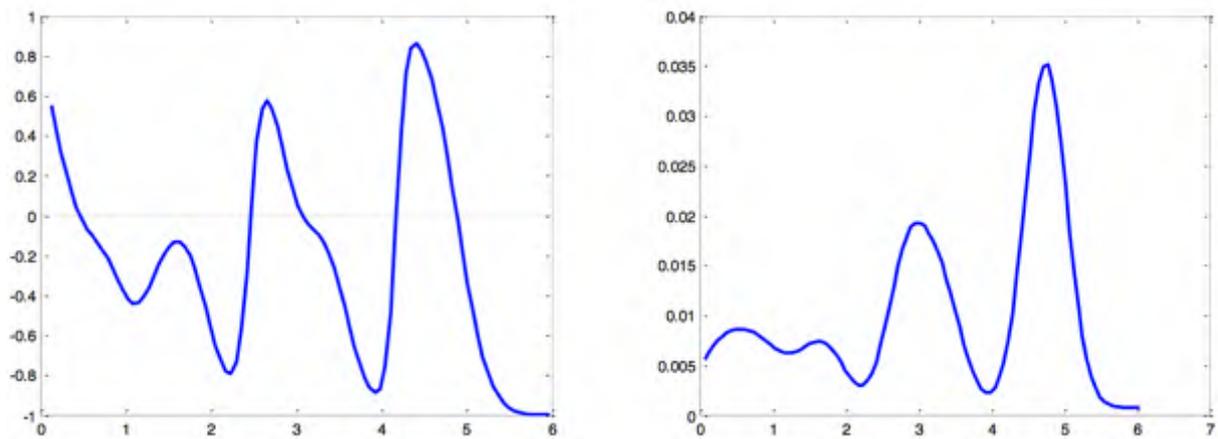
**Figure** 11 The size distribution dynamics of the western cities for the period 1984-2010

## 6. Conclusion and discussion

The issues whether city size growth parallel, converge or diverge have attracted many attentions in urban studies. This paper examines the size distribution dynamics of Chinese cities during the period 1984-2010. It differs from existing studies in using a new framework named distribution dynamics approach, which can provide dynamic behavior of the entire distribution shape of city size. In particular, this approach has advantages in analyzing persistence and convergence clubs. The results in this paper show that persistence and multimodality are the dominant characteristics in the distribution dynamics of the Chinese cities. Some studies use persistence as the evidence of the paralleled growth of city size. However, the multimodality found in long run steady distribution for all cities and regional city groups implies that this may be misleading. The distribution dynamics of city size may be more complicated than one expected. New models should be developed to shed light on this new stylized fact. Our results show that Chinese cities exhibit significant convergence in the sample period, 61.8 percent of cities have tendency to move down in the intra-distribution dynamics.

Similar to that predicted by the theory of sequential growth, our results show that initial city size may have important impact on the following distribution dynamics of city size. Initially small and medium-sized cities exhibit strong tendency of convergence, while large cities show significant persistence and multimodality in our sample period. This result does not support the sequential growth theory that large cities have more mobility than small- and medium-sized cities. The path dependence effect, if any, would be stronger in the large cities than in small- and medium-sized cities. This result shows that examinations focusing only on the upper tail large cities, such as most empirical studies on the Zipf's law, may neglect important information on the growth patterns of city size. The city size distribution also differs in several aspects among various regions. There are more megacities in the eastern region than in the central and western region in China. Moreover, the central and western cities have stronger tendency to move downward in the intra-distribution dynamics than the eastern cities do. This implies the relative shrinkage of city size in the central and western regions. Since the Economic Reform and the Opening Policy in 1980s, the central and western regions have been suffering from the loss of population relative to the eastern regions. Our findings suggest that locational fundamentals may have impact on the distribution dynamics of city size. Future research should focus on how location-specific factors, such as weather conditions, natural resources, and governmental regional development programs, affect distribution dynamics of city size.

The findings in this paper have important policy implication. China is experiencing rapid urbanization. The Chinese government play important role in this process through investment policy. Most of the urban investment is extremely durable. Knowing the distribution dynamics of city size in the long run may help the government to make reasonable urban planning and improve economic efficiency in the urbanization process.

**Acknowledgements**

Work of this paper was supported by the China National Social Science Foundation (No. 15ZDA054), Humanities and Social Sciences of Ministry of Education Planning Fund (16YJA790050), and the National Natural Science Foundation of China ((Nos. 71473105, 71273261 and 71573258).

**Reference**

[1] G. H. Overman, M. Y. Ioannides, Cross-sectional evolution of the US city size distribution. Journal of Urban Economics. 49(3) (2001) 543-566.

[2] R. González-Val, A. Ramos, F. Sanz-Gracia, M. Vera-Cabello, Size distributions for all cities: Which one is best? Papers in Regional Science. 94(1) (2015)177-196.

[3] E. Rossi-Hansberg, L. M. Wright, Urban structure and growth. The Review of Economic Studies. 74(2) (2007)597-624.

[4] Z. Chen, S. Fu, D. Zhang, Searching for the parallel growth of cities in China. Urban Studies. 50(10) (2013)2118-2135.

[5] T. D. Quah, Twin peaks: growth and convergence in models of distribution dynamics. The Economic Journal. 106(437) (1996)1045-1055.

[6] J. Eaton, Z. Eckstein, Cities and growth: Theory and evidence from France and Japan. Regional science and urban Economics. 27(4) (1997) 443-474.

[7] G. Anderson, Y. Ge, The size distribution of Chinese cities. Regional Science and Urban Economics. 35(6) (2005)756-776.

[8] Z. Xu, N. Zhu, City size distribution in china: are large cities dominant? Urban Studies, 46(10) (2009) 2159-2185.

[9] D. Black, Henderson V., Urban evolution in the USA. Journal of Economic Geography. 3(4) (2003)343-372.

[10] M.Y. Ioannides, H. G. Overman, Zipf's Law for cities: An Empirical examination. Regional Science and Urban Economics. 33 (2003)127-137.

[11] D. Black, V. Henderson, A theory of urban growth. Journal of Political Economy. 107(2) (1999) 252-284.

[12] T. K. Soo, Zipf's Law for cities: a cross-country investigation. Regional Science and Urban Economics. 35(3) (2005) 239-263.

[13] G. Duranton, Urban evolutions: the fast, the slow, and the still. American Economic Review. 97 (1) (2007)197–221.

[14] M. Bosker, S. Brakman, H. Garretsen, M. Schramm, A century of shocks: the evolution of the German city size distribution 1925–1999.Regional Science and Urban Economics. 38(4) (2008)330-347.

[15] T. K. Soo, Zipf, Gibrat and geography: evidence from China, India and Brazil. Papers in Regional Science. 93(1) (2014)159-181.

[16] C. L. Malacarne, S. R. Mendes, K. E. Lenzi, q-Exponential distribution in urban agglomeration. Physical Review E, 65(1) (2001) 017106.

[17] T. K. Soo, Zipf's Law and urban growth in Malaysia. Urban Studies. 44(1) (2007)1-14.

[18] W. Reed, On the rank-size distribution for human settlements. Journal of Regional Science. 42(2002)1–17.

[19] K. Giesen, J. Südekum, Zipf's law for cities in the regions and the country. Journal of Economic Geography. 68(2) (2010)129-137.

[20] Y. Ioannides, S. Skouras, US city size distribution: Robustly Pareto, but only in the tail. Journal of Urban Economics. 73(1) (2013)18-29.

[21] X. Gabaix, Zipf's law and the growth of cities. American Economic Review. 43(89) (1999)129-132.

[22] C. J. Córdoba, On the distribution of city sizes. Journal of Urban Economics. 63(1) (2008)177-197.

[23] S. Sharma, Persistence and stability in city growth. Journal of Urban Economics. 53(2) (2003) 300-320.

[24] Z. Wang, J. Zhu, Evolution of China's city-size distribution: empirical evidence from 1949 to 2008. Chinese Economy. 46(1) (2013) 38-54.

[25] V. J. Henderson, J. A. Venables, Dynamics of city formation. Review of Economic Dynamics. (2) (2009)233–254.

[26] D. Cuberes, A model of sequential city growth. The BE Journal of Macroeconomics. 9(1) (2009)1-41.

[27] D. Cuberes, Sequential city growth: Empirical evidence. Journal of Urban Economics. 69(2) (2011) 229-239.

[28] M. Sánchez-Vidal, R. González-Val, E. Viladecans-Marsal, Sequential city growth in the US: Does age matter? Regional Science and Urban Economics. 44(2014) 29-37.

[29] K. Sheng, W. Sun, J. Fan, Sequential city growth: Theory and evidence from the US. Journal of Geographical Sciences. 24(6) (2014)1161-1174.

[30] T. D. Quah, Empirical cross-section dynamics in economic growth. European Economic Review, 37(2) (1993)426-434.

[31] T. D. Quah, Empirics for growth and distribution: stratification, polarization, and convergence clubs. Journal of Economic Growth. 2(1) (1997)27-59.

[32] S. Bulli, Distribution dynamics and cross-country convergence: A new approach. Scottish Journal of Political Economy. 48(2) (2001) 226-243.

[33] A. P. Johnson, A Continuous State Space Approach to "Convergence by Parts". Economics Letters. 86(3) (2005)317-321.

[34] F. Juessen, A distribution dynamics approach to regional GDP convergence in unified Germany. Empirical Economics, 37(3) (2009)627-652.

[35] W. B. Silverman, Density estimation for statistics and data analysis, CRC press. 1986.

[36] S. T. Cheong, Y. Wu, Convergence and transitional dynamics of china's industrial output: a county-level study using a new framework of distribution dynamics analysis. University of Western Australia, Business School DISCUSSION PAPER 14.21, 2014.

[37] NBSC (National Bureau of Statistics of China), China City Statistical Yearbook. Beijing: China Statistic Press. (2003-2012a).

[38] M. Fujita, T. Mori, Structural stability and evolution of urban systems. Regional Science & Urban Economics. 27(4–5) (1997) 399-442.

[39] J. L. Fan, H. Liao, B. J. Tang, S. Y. Pan, H. Yu, Y. M. Wei, The impacts of migrant workers consumption on energy use and $CO_2$ emissions in China. Natural Hazards, 81(2) (2016)725-743.